\documentclass[prd,aps,groupedaddress,showpacs,amsmath]{revtex4}
\usepackage{epsf,epsfig,graphicx}

\newcounter{muni}

\usepackage{graphicx}

\newcommand{\etaq}{\eta_{q}}
\newcommand{\etas}{\eta_{s}}

\begin{document}
\hbadness=10000 \pagenumbering{arabic}

\title{Okubo-Zweig-Iizuka-rule violation and $B\to \eta^{(\prime)}K$ branching ratios}

\author{Jen-Feng Hsu$^{1}$}
\email{kleinhsu@gmail.com}
\author{Yeo-Yie Charng$^{1}$}
\email{charng@phys.sinica.edu.tw}
\author{Hsiang-nan Li$^{1,2,3}$}
\email{hnli@phys.sinica.edu.tw}

\affiliation{$^{1}$Institute of Physics, Academia Sinica, Taipei,
Taiwan 115, Republic of China} \affiliation{$^{2}$Department of
Physics, National Cheng-Kung University, Tainan, Taiwan 701,
Republic of China} \affiliation{$^{3}$Department of Physics,
National Tsing-Hua University, Hsinchu, Taiwan 300, Republic of
China}

\begin{abstract}

We show that few-percent Okubo-Zweig-Iizuka-rule violating effects
in the quark-flavor basis for the $\eta$-$\eta'$ mixing can enhance
the chiral scale associated with the $\eta_q$ meson few times. This
enhancement is sufficient for accommodating the dramatically
different data of the $B\to\eta^{\prime} K$ and $B\to\eta K$
branching ratios. We comment on other proposals for resolving this
problem, including flavor-singlet contributions, axial $U(1)$
anomaly, and nonperturbative charming penguins. Discrimination of
the above proposals by means of the $B\to\eta^{(\prime)}\ell\nu$ and
$B_s\to\eta^{(\prime)}\ell\ell$ data is suggested.

\end{abstract}


\maketitle

\section{INTRODUCTION}

The large $B\to\eta^{\prime} K$ and small $B\to\eta K$ branching
ratios measured by the $B$ factories are still not completely
understood \cite{HFAG}:
\begin{eqnarray}
B(B^\pm\to\eta' K^\pm)&=&(70.2\pm 2.5)\times
10^{-6}\;,\nonumber\\
B(B^0\to\eta' K^0)&=&(64.9\pm 3.1)\times
10^{-6}\;,\nonumber\\
B(B^\pm\to\eta K^\pm)&=&(2.7\pm 0.3)\times 10^{-6}\;,\nonumber\\
B(B^0\to\eta K^0)&<&1.9\times 10^{-6}\;.\label{data}
\end{eqnarray}
The predictions for $B(B\to\eta' K)$ from both the perturbative QCD
(PQCD) \cite{KS01} and QCD-improved factorization (QCDF) \cite{BN02}
approaches in the Feldmann-Kroll-Stech (FKS) scheme \cite{FKS}
for the $\eta$-$\eta'$ mixing are smaller than the data. Several
resolutions to this puzzle have been proposed: a significant
flavor-singlet contribution \cite{BN02}, a large $B\to\eta'$
transition form factor \cite{TNP07}, a high chiral scale $m_0^q$
\cite{ACG07} associated with the $\eta_q$ meson which is composed of
the $u\bar u$ and $d\bar d$ content in the quark-flavor basis
\cite{FKS}, an enhanced hadronic matrix element $\langle
0|\bar{s}\gamma_{5}s|\eta'\rangle$ \cite{GK06} of the strange-quark
pseudoscalar density due to axial $U(1)$ anomaly \cite{GT04}, the
long-distance charming penguin and gluonic charming penguin
\cite{WZ0610} in the soft-collinear effective theory (SCET)
\cite{BPRS,Chay:2003ju}, and inelastic final-state interaction (FSI)
\cite{CCS05}. The motivation of \cite{CCS05} is to fix FSI effects
using the data in Eq.~(\ref{data}), and then to predict CP
asymmetries in the $B\to\eta^{(\prime)}K$ decays.

A sizable gluonic content in the $\eta'$ meson was indicated from a
phenomenological analysis of the relevant data \cite{K99}, and also
by the recent KLOE measurement \cite{KLOE} (but see an opposite
observation in \cite{RE07}). The flavor-singlet contributions to the
$B\to\eta^{(\prime)} K$ branching ratios, containing those from the
$b\to sgg$ transition \cite{EKP04}, from the spectator scattering
\cite{Du:1997hs,DYZ99}, and from the weak annihilation, have been
taken into account in QCDF \cite{BN02}. However, the gluonic
contribution to the $B\to\eta'$ transition form factor was
parameterized and increased arbitrarily up to 40\% \cite{BN02} in
order to explain Eq.~(\ref{data}). This piece was also included in
the parametrization for two-body nonleptonic $B$ meson decay
amplitudes based on SCET, but found to be destructive to the quark
contribution from data fitting \cite{WZ0610}.  To settle down this
issue, we have examined the gluonic contribution in the PQCD
approach \cite{LY1,KLS,LUY,LK04} with the associated parameters
being experimentally constrained, and observed that it is
constructive and negligible (of few percents at most) in the
$B\to\eta^{(\prime)}$ transitions \cite{Charng}. Our conclusion has
been confirmed by the sum-rule analysis in \cite{BJ07}. If so, one
has to clarify what mechanism is responsible for the increase of the
$B\to\eta'$ form factor postulated in \cite{TNP07}.

The chiral scale for the $\eta_q$ meson is defined by $m_0^q\equiv
m_{qq}^2/(2m_q)$ with the light quark mass $m_q=m_u=m_d$ under the
exact isospin symmetry. The mass $m_{qq}$ was increased from its
generally accepted value 0.11 GeV, close to the pion mass, to $0.22$
GeV in \cite{ACG07}. This enhancement then gives a larger
$B\to\eta_q K$ decay amplitude, a more destructive (constructive)
interference with the $B\to\eta_s K$ amplitude \cite{HJL91}, where
the $\eta_s$ meson is composed of the $s\bar s$ content in the
quark-flavor basis, and thus a smaller $B\to\eta K$ (larger
$B\to\eta'K$) branching ratio. It has been found that the PQCD
results for the $B\to\eta^{(\prime)}K$ branching ratios
corresponding to $m_{qq}=0.22$ GeV agree with the data \cite{ACG07}.
Note that the PQCD results for the $B\to\eta^{(\prime)}K^*$
branching ratios are also consistent with the data, which show a
tendency opposite to Eq.~(\ref{data}): $B(B^\pm\to\eta' K^{*})$ are
smaller than $B(B^\pm\to\eta K^*)$ by about a factor 4 \cite{HFAG}.
Whether there is any mechanism to achieve the enhancement of
$m_{qq}$ is not clear. We shall argue that a tiny effect violating
the Okubo-Zweig-Iizuka (OZI) rule \cite{OZI}, which was neglected in
the FKS scheme, could be the responsible mechanism.

The OZI-rule violation has been studied in, for example, exclusive
$\eta^{(\prime)}$ productions from $\pi N$ and $NN$ scattering in a
wide range of energy scales \cite{NS03}. Most of the observed ratios
of the cross sections, $\sigma(\pi N, NN\to\eta X)/\sigma(\pi
N,NN\to\eta' X)$, are in agreement with or slightly larger than the
expectation around 1.5 from the FKS scheme, considering experimental
uncertainties. The exceptions with significant OZI-rule violation
appear in the $\eta^{(\prime)}$ productions at thresholds: the ratio
$\sigma(pp\to pp\eta)/\sigma(pp\to pp\eta')$ was measured to be
$37.0\pm 11.3$ and $26.2\pm 5.4$ with the proton energy being 2.9
MeV and 4.1 MeV, respectively \cite{NS03,PM00}. The above tendency
hints the possibility of small OZI-rule violation in $B$ and $D$
meson decays into light final states, whose energy release is of
order few GeV. The proposal in \cite{GK06} relies on the large
matrix element $\langle 0|\bar{s}\gamma_{5}s|\eta'\rangle$, which
strengthens its difference from $\langle
0|\bar{s}\gamma_{5}s|\eta\rangle$, and the difference between the
$B\to\eta' K$ and $B\to\eta K$ branching ratios through penguin
contributions. It will be pointed out that this proposal demands
larger OZI-rule violation, which is not obviously signaled in the
$D_s\to\eta^{(\prime)}\ell\nu$ data.

In Sec.~II we show that few-percent OZI violating effects enhance
the mass $m_{qq}$ sufficiently, which accommodates the data of the
$B\to\eta^{(\prime)}K$ branching ratios in the PQCD approach. In
Sec.~III we make a critical review on other proposals for this
subject from both theoretical and experimental points of view.
Section~IV contains a summary, in which experimental discrimination
for all the proposed mechanisms is suggested.

\section{OZI-RULE VIOLATION}

We consider the following OZI-rule violating matrix elements in the
quark-flavor basis,
\begin{eqnarray}
\left\langle0\left|\bar{q}\gamma^{\mu}\gamma_{5}q\right|\eta_{s}(P)
\right\rangle&=&\frac{i}{\sqrt{2}}f_{qs}P^{\mu}\;,\nonumber\\
\left\langle0\left|\bar{s}\gamma^{\mu}\gamma_{5}s\right|\etaq(P)
\right\rangle&=&if_{sq}P^{\mu}\;,
\end{eqnarray}
for the light quark $q=u$ or $d$, where the decay constants
$f_{qs}$ and $f_{sq}$ are expected to be small and have been
neglected in the FKS scheme. We also define the decay constants
for the $\eta_{q,s}$ mesons and for the $\eta^{(\prime)}$ mesons:
\begin{eqnarray}
\left\langle0\left|\bar{q}\gamma^{\mu}\gamma_{5}q\right|\etaq(P)
\right\rangle&=&\frac{i}{\sqrt{2}}f_{qq}P^{\mu}\;,\nonumber\\
\left\langle0\left|\bar{s}\gamma^{\mu}\gamma_{5}s\right|\eta_{s}(P)
\right\rangle&=&if_{ss}P^{\mu}\;, \nonumber\\
   \langle 0|\bar q\gamma^\mu\gamma_5 q|\eta^{(\prime)}(P)\rangle
   &=& \frac{i}{\sqrt2}\,f_{\eta^{(\prime)}}^q\,P^\mu \;,\nonumber \\
   \langle 0|\bar s\gamma^\mu\gamma_5 s|\eta^{(\prime)}(P)\rangle
   &=& i f_{\eta^{(\prime)}}^s\,P^\mu \;.\label{deffh}
\end{eqnarray}
The physical states $\eta$ and $\eta'$ are related to the flavor
states $\eta_q$ and $\eta_s$ through
\begin{equation}\label{qs}
   \left( \begin{array}{c}
    |\eta\rangle \\ |\eta'\rangle
   \end{array} \right)
   = U(\phi)
   \left( \begin{array}{c}
    |\eta_q\rangle \\ |\eta_s\rangle
   \end{array} \right) \;,
\end{equation}
with the unitary matrix
\begin{equation}
U(\phi)=\left( \begin{array}{cc}
    \cos\phi & ~-\sin\phi \\
    \sin\phi & \phantom{~-}\cos\phi
   \end{array} \right)\;.
\end{equation}
The above decay constants are transformed into each other via
\begin{equation}
    \left(
      \begin{array}{clrr}
        f^{q}_{\eta} & f^{s}_{\eta}  \\
        f^{q}_{\eta^{'}} & f^{s}_{\eta^{'}}
    \end{array}
    \right)
    =U(\phi)
    \left(
      \begin{array}{clrr}
        f_{qq} & f_{sq}  \\
        f_{qs} & f_{ss}  \\
    \end{array}
    \right)\;.\label{ori}
\end{equation}

We repeat the derivation of Eq.~(7) in \cite{Charng}, obtaining
\begin{equation}
  M^{2}_{qs}=U^{\dagger}(\phi)M^{2}U(\phi)
  \left(
  \begin{array}{ccrr}
      1 & Y_{sq} \\
      Y_{qs} & 1 \\
  \end{array}
 \right)\;,\label{m2qs}
\end{equation}
where the OZI violating parameters are defined by $Y_{qs}\equiv
f_{qs}/f_{qq}$ and $Y_{sq}\equiv f_{sq}/f_{ss}$, and the mass
matrices written as
\begin{eqnarray}
M^2&=&\left(\begin{array}{cc}
  m_{\eta}^2 & 0 \\
  0 & m_{\eta'}^2 \\
\end{array} \right)\;,\nonumber\\
M_{qs}^2&=&\left(\begin{array}{cc}
m_{qq}^2+(\sqrt{2}/f_{qq})\langle 0|\alpha_sG{\tilde
G}/(4\pi)|\eta_q\rangle &  (1/f_{ss})\langle
0|\alpha_sG{\tilde G}/(4\pi)|\eta_q\rangle \\
               (\sqrt{2}/f_{qq})\langle
0|\alpha_sG{\tilde G}/(4\pi)|\eta_s\rangle &
m_{ss}^2+(1/f_{ss})\langle 0|\alpha_sG{\tilde
G}/(4\pi)|\eta_s\rangle
\\
\end{array}\right)\;,\label{mmqs}
\end{eqnarray}
with the abbreviations
\begin{eqnarray}
m_{qq}^2&=&\frac{\sqrt{2}}{f_{qq}}\langle 0|m_u\bar u i\gamma_5
u+m_d\bar d
i\gamma_5 d|\eta_q\rangle\;,\nonumber\\
m_{ss}^2&=&\frac{2}{f_{ss}}\langle 0|m_s\bar s i\gamma_5
s|\eta_s\rangle\;.\label{mqq}
\end{eqnarray}
Note that the matrix $M_{qs}^2$ becomes non-hermitian, after
including the OZI violating effects, or employing the two-angle
mixing formalism [see Eq.~(\ref{alt}) below]. In fact, this matrix
is hermitian only in the FKS scheme.

Equation~(\ref{m2qs}) determines the four elements in $M^2_{qs}$:
\begin{eqnarray}
m_{qq}^{2}&=&m_{qq}^{(0)2}+\left[
Y_{qs}(m_{\eta'}^2-m_\eta^2)\cos\phi\sin\phi-\frac{\sqrt{2}f_{ss}}{f_{qq}}Y_{sq}
(m_{\eta}^2\cos^2\phi+m_{\eta'}^2\sin^2\phi)
\right]\;,\nonumber\\
m_{ss}^{2}&=&m_{ss}^{(0)2}+
\left[Y_{sq}(m_{\eta'}^2-m_\eta^2)\cos\phi\sin\phi
-\frac{f_{qq}}{\sqrt{2}f_{ss}}Y_{qs}(m_{\eta'}^2\cos^2\phi+m_{\eta}^2\sin^2\phi)
\right]\;,\label{chiral}
\end{eqnarray}
with the original solutions \cite{FKS}
\begin{eqnarray}
m_{qq}^{(0)2}&=&m_{\eta}^2\cos^2\phi+m_{\eta'}^2\sin^2\phi-
\frac{\sqrt{2}f_{ss}}{f_{qq}}(m_{\eta'}^2-m_\eta^2)\cos\phi\sin\phi\;,\label{mqq0}\\
m_{ss}^{(0)2}&=&m_{\eta'}^2\cos^2\phi+m_{\eta}^2\sin^2\phi-
\frac{f_{qq}}{\sqrt{2}f_{ss}}(m_{\eta'}^2-m_\eta^2)\cos\phi\sin\phi\;.
\end{eqnarray}
Substituting the parameters extracted in \cite{FKS}
\begin{eqnarray}
& &f_{q}=(1.07\pm 0.02)f_\pi\;,\;\;\;\;f_{s}=(1.34\pm
0.06)f_\pi\;,\;\;\; \phi=39.3^\circ\pm 1.0^\circ\;, \label{qsp}
\end{eqnarray}
for $f_{qq}$, $f_{ss}$ and $\phi$ in Eq.~(\ref{mqq0}), respectively,
and adopting the masses $m_\eta=0.548$ GeV and $m_{\eta'}=0.958$
GeV, we derive
\begin{eqnarray}
m_{qq}^{(0)}\approx 0.11\;\;{\rm GeV}\;,\;\;\;\;m_{ss}^{(0)}\approx
0.71\;\;{\rm GeV}\;.\label{sol}
\end{eqnarray}
The smallness of $m_{qq}^{(0)}$ is attributed to the strong
cancelation between the two terms on the right-hand side of
Eq.~(\ref{mqq0}), where the second term is associated with the axial
$U(1)$ anomaly. It is then expected that $m_{qq}^{(0)}$ is easily
affected by the OZI violating contribution, while the larger
$m_{ss}^{(0)}$ is stable. If stretching the uncertainties of $f_q$,
$f_s$, and $\phi$ in Eq.~(\ref{qsp}), $m_{qq}^{(0)}$ could reach
0.22 GeV without the OZI violating effect. Nevertheless, the ranges
of these parameters depend on data included for fit (different sets
of data lead to different ranges), and on theoretical modelling of
considered processes \cite{Feld99}. Here we suggest a plausible
mechanism, which easily modifies $m_{qq}$ without stretching
uncertainties.

\begin{figure}[ht]
\begin{center}
\resizebox{7cm}{!}{\includegraphics{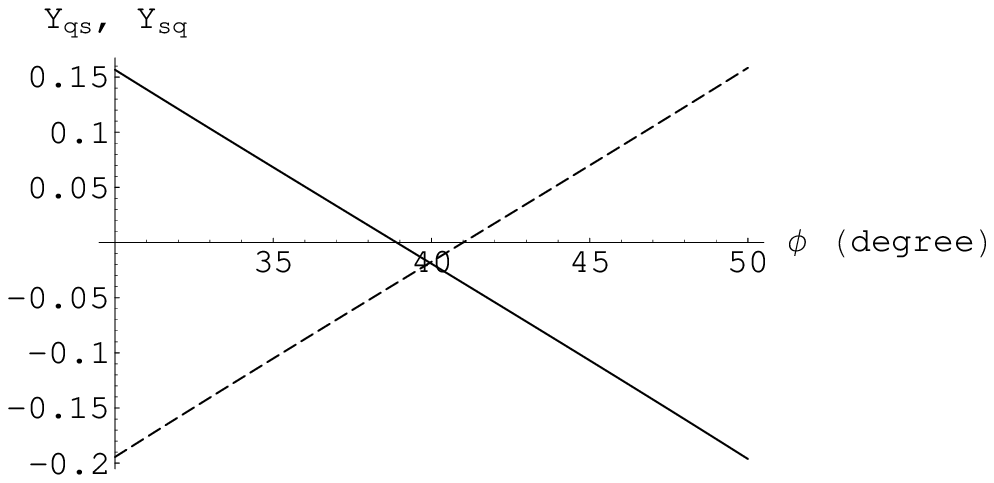}} \ \hskip1cm
\resizebox{7cm}{!}{\includegraphics{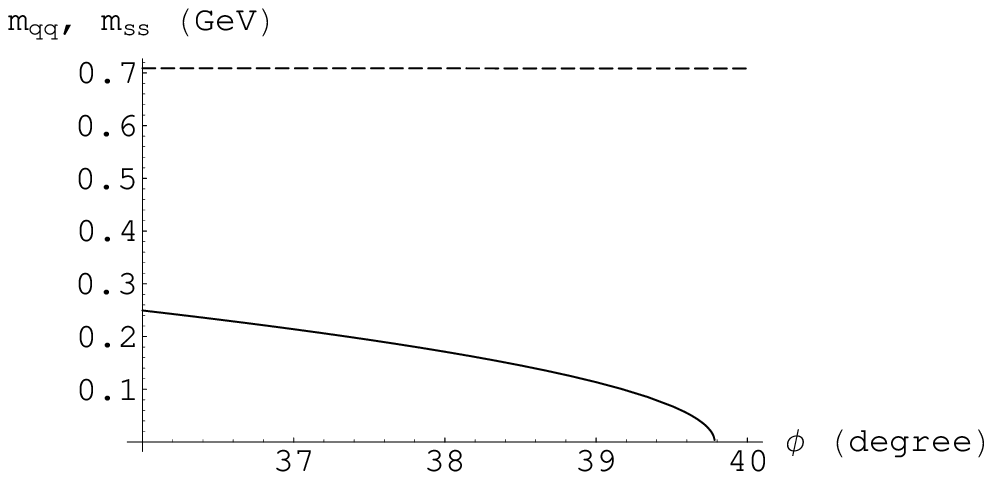}} \ \

(a)\hskip8cm (b)

\caption{Dependences of (a) $Y_{qs}$ (solid line) and $Y_{sq}$
(dashed line) and of (b) $m_{qq}$ (solid line) and $m_{ss}$
(dashed line) on $\phi$.}\label{figm0}
\end{center}
\end{figure}

The order of magnitude of $f_{qs,sq}$ can be estimated via the
two-angle mixing formalism \cite{SSW,HL98}
\begin{equation}
    \left(
  \begin{array}{clrr}
      f^{q}_{\eta} & f^{s}_{\eta} \\
      f^{q}_{\eta^{'}} & f^{s}_{\eta^{'}} \\
  \end{array}
 \right)
 =U_{qs}
  \left(
  \begin{array}{clrr}
      f_{q} & 0 \\
      0 & f_{s} \\
  \end{array}
 \right)\;,\label{alt}
\end{equation}
with the matrix
\begin{equation}
    U_{qs}\equiv
    \left(
  \begin{array}{cc}
      \cos{\phi_{q}} & -\sin\phi_{s} \\
      \sin\phi_{q} & \cos\phi_{s} \\
  \end{array}
 \right)\;.
\end{equation}
If $\phi_q=\phi_s$, the above formalism reduces to the FKS scheme;
that is, the OZI violating matrix elements give rise to the
difference between $\phi_q$ and $\phi_s$, or to the energy
dependence of the mixing angle introduced in \cite{Escribano99}. We
insert a typical set of parameters \cite{Escribano99},
\begin{equation}
  \begin{array}{clrr}
  f_q=1.10f_{\pi}\;,\;\;\;\;f_s=1.46f_{\pi}\;,\;\;\;\;
  \phi_{q}=38.9^{\circ}\;,\;\;\;\; \phi_{s}=41.0^\circ\;,
  \end{array}\label{two}
\end{equation}
into Eq.~(\ref{alt}), compute the left-hand side of Eq.~(\ref{alt}),
and then invert Eq.~(\ref{ori}) to obtain the $\phi$ dependence of
the OZI violating parameters $Y_{qs,sq}$. The $\phi$ dependences of
$Y_{qs,sq}$ and of $m_{qq,ss}$ from Eq.~(\ref{chiral}) in a
reasonable range of $\phi$, roughly from $33^\circ$ to $42^\circ$
\cite{CET07,LD07,HW06,RE08}, are displayed in Fig.~\ref{figm0}. It
indicates that the tiny $Y_{qs}=0.036$ and $Y_{sq}=-0.073$ (for
$\phi=36.84^\circ$) reproduce the inputs in \cite{ACG07},
\begin{eqnarray}
m_{qq}= 0.22\;\;{\rm GeV}\;,\;\;\;\; m_{ss}= 0.71 \;\;{\rm
GeV}\;,\label{res}
\end{eqnarray}
namely, give a factor-2 enhancement of $m_{qq}$, and almost no
impact on $m_{ss}$.

An updated fitting leads to similar results but with a higher
$f_s\approx 1.66f_\pi$ compared to that in Eq.~(\ref{two}), which is
mainly attributed to the change of the $\phi\to\eta'\gamma$ data
\cite{Escribano}. In this case $m_{qq}$ reaches about 0.2 GeV for a
lower value of $\phi\approx 34^\circ$. In general, we should
introduce the additional OZI violating matrix elements into
Eq.~(\ref{mmqs}),
\begin{eqnarray}
m^{2}_{qs}&=&\frac{\sqrt{2}}{f_{qq}}\left\langle0
\left|m_{u}\bar{u}i\gamma_{5}u+m_{d}\bar{d}i\gamma_{5}d
\right|\etas\right\rangle\;,\nonumber\\
m^{2}_{sq}&=&\frac{2}{f_{ss}}\left\langle0
\left|m_{s}\bar{s}i\gamma_{5}s\right|\etaq\right\rangle\;,\label{mqs}
\end{eqnarray}
whose inclusion, however, modifies $m_{qq}$ and $m_{ss}$ only
slightly as shown later. Besides, the isospin breaking effect from
mixing with pions is also negligible. This effect, found to be of
few percents \cite{K05}, appears quadratically in the expressions of
$m_{qq}^2$, namely, at the $10^{-4}$ level.


Corresponding to Eq.~(\ref{res}), the $B\to\eta^{(\prime)}K$
branching ratios were found to be
\begin{eqnarray}
B(B^\pm\to\eta' K^\pm)&=&65.04 (34.60) \times
10^{-6}\;,\nonumber\\
B(B^0\to\eta' K^0)&=&62.69 (31.44) \times
10^{-6}\;,\nonumber\\
B(B^\pm\to\eta K^\pm)&=&1.52 (5.66) \times 10^{-6}\;,\nonumber\\
B(B^0\to\eta K^0)&=&1.43 (3.01) \times 10^{-6}\;,
\end{eqnarray}
in the PQCD approach \cite{ACG07}, where the results for
$m_{qq}=0.14$ GeV are quoted in the parentheses for comparison.
Obviously, the agreement with the data in Eq.~(\ref{data}) has been
greatly improved. Note that our point is not to claim the existence
of the OZI violating effects in the $B\to\eta^{(\prime)}K$ decays,
but that just few percents of such effects, which are very likely
viewing the data of other $\eta^{(\prime)}$ involved processes
\cite{NS03}, are sufficient for resolving the puzzle.

The consistency of the PQCD results \cite{ACG07} with the data of
the $B\to\eta^{(\prime)}K^*$ branching ratios is also improved by
increasing $m_{qq}$. The data \cite{HFAG}
\begin{eqnarray}
B(B^\pm\to\eta' K^{*\pm})&=&(4.9^{+2.1}_{-1.9})\times
10^{-6}\;,\nonumber\\
B(B^0\to\eta' K^{*0})&=&(3.8\pm 1.2)\times
10^{-6}\;,\nonumber\\
B(B^\pm\to\eta K^{*\pm})&=&(19.3\pm 1.6)\times 10^{-6}\;,\nonumber\\
B(B^0\to\eta K^{*0})&=&(15.9\pm 1.0)\times 10^{-6}\;,\label{dats}
\end{eqnarray}
exhibit a tendency opposite to that of the $B\to\eta^{(\prime)}K$
branching ratios in Eq.~(\ref{data}), which is attributed to the
sign flip of the $(V-A)(V+A)$ penguin contribution in the
$B\to\eta_sK^*$ decays (involving the $B\to K^*$ transition form
factor) \cite{BN02,LM06}, i.e., to an opposite interference pattern
between the $B\to\eta_qK^*$ and $B\to\eta_sK^*$ amplitudes.
Similarly, it is difficult to accommodate the factor-4 difference
between the measured $B\to\eta'K^*$ and $B\to\eta K^*$ branching
ratios in Eq.~(\ref{dats}) in the FKS scheme. Additional mechanism,
such as a significant flavor-singlet contribution \cite{BN02} or a
$B\to\eta_qK^*$ decay amplitude enhanced by a large $m_{qq}$
\cite{ACG07}, is required. We mention that the absolute
$B\to\eta^{(\prime)}K^*$ branching ratios predicted in \cite{BN02}
in the default scenario are smaller than the data in
Eq.~(\ref{dats}), which is a general trend of the QCDF approach to
$B\to PV$ decays \cite{BN03}, with $P$ ($V$) denoting a pseudoscalar
(vector) meson.

\begin{table}[ht]
\begin{center}
\begin{tabular}{ccccc}
\hline\hline
form factors          & $F^{B\eta}_{+,0}$ & $F^{B\eta}_T$ & $F^{B\eta^\prime}_{+,0}$ & $F^{B\eta^\prime}_T$ \\
\hline
  $F(0)$               & 0.308            & 0.298         & 0.235                   & 0.227  \\
\hline
  $\phi_{qs}^A$ contribution (\%)  &-0.386 & -0.330       & 0.673              & 0.577  \\
\hline
  gluonic contribution (\%)        & 0.196 & 0.169         & 1.24              & 1.07  \\
\hline\hline
\end{tabular}
\caption{\label{tab} Contributions to the $B\to\eta^{(\prime)}$ form
factors at maximal recoil from the distribution amplitudes in
Eq.~(\ref{phiqs}) and from the gluonic content for $f_{qs}=5.14$
MeV, $\phi=36.84^\circ$, and $m_{qq}=0.22$ GeV.}
\end{center}
\end{table}

The introduction of the OZI violating decay constants $f_{qs,sq}$
implies the additional twist-2 $\eta_{q,s}$ meson distribution
amplitudes,
\begin{eqnarray}
\left\langle\eta_s(P)\left|\bar{q}^{a}_{\gamma}(z)q^{b}_{\beta}(0)
\right|0\right\rangle&=&-\frac{i}{2\sqrt{N_c}} \delta^{ab}\int_0^1dx
e^{ixP\cdot z}[\gamma_5\not P]_{\beta\gamma}\phi^A_{qs}(x)\;,
\nonumber\\
\left\langle\eta_{q}(P)\left|\bar{s}^{a}_{\gamma}(z)s^{b}_{\beta}(0)
\right|0\right\rangle&=&-\frac{i}{\sqrt{2N_c}} \delta^{ab}\int_0^1dx
e^{ixP\cdot z}[\gamma_5\not
P]_{\beta\gamma}\phi^A_{sq}(x\;)\;.\label{phiqs}
\end{eqnarray}
We show that these distribution amplitudes need not to be considered
by taking the semileptonic decays $B\to\eta^{(\prime)}\ell\nu$ as an
example. $\phi^A_{sq}$ is irrelevant at the current level of
accuracy, since it contributes at next-to-leading order in
$\alpha_s$: it is involved in the diagram, where the light-quark
pair from the $B$ meson transition converts into a pair of valence
strange quarks in the $\eta_q$ meson through two-gluon exchanges.
Therefore, we only examine the contribution from $\phi^A_{qs}$ to
the $B\to\eta^{(\prime)}$ transition form factors $F_{+,0,T}$
defined via the matrix elements,
\begin{eqnarray}
\langle \eta^{(\prime)}(P_2)|{\bar b}\gamma_\mu u|B(P_1)\rangle
&=&F_{+}^{B\eta^{(\prime)}}(q^2)\left[(P_1+P_2)_\mu
-\frac{m_B^2-m_{\eta^{(\prime)}}^2}{q^2}q_\mu\right]\nonumber\\
& &+F_{0}^{B\eta^{(\prime)}}(q^2)
\frac{m_B^2-m_{\eta^{(\prime)}}^2}{q^2}q_\mu\;,\nonumber\\
\langle \eta^{(\prime)}(P_2)|{\bar b} i\sigma^{\mu\nu} q_\nu
u|B(P_1) \rangle &=& \frac{
F_T^{B\eta^{(\prime)}}(q^2)}{m_B+m_{\eta^{(\prime)}}}\left[
(m_B^2-m_{\eta^{(\prime)}}^2)\,q^\mu-q^2(P_1^\mu+P_2^{\mu})\right]\;,
\label{ftensor}
\end{eqnarray}
with the $B$ meson momentum $P_1$, the $\eta^{(\prime)}$ meson
momentum $P_2$, and the lepton-pair momentum $q=P_1-P_2$.

The corresponding PQCD factorization formulas are referred to
\cite{Charng}, and the Gegenbauer moments for the models of the
$\eta_{q}$ meson distribution amplitudes are the same as in
\cite{ACG07,WLX05}. We also compute the gluonic contribution for
comparison \footnote{The formulas for the gluonic contribution in
Phys. Rev. D {\bf 74}, 074024 (2006) \cite{Charng} should be
multiplied by a factor $1/4N_c$ as corrected in
arXiv:hep-ph/0609165.}.  Assuming the asymptotic form
$\phi^A_{qs}(x)=3f_{qs}x(1-x)/\sqrt{6}$, the numerical results of
the form factors $F_{+,0,T}^{B\eta}$ and $F_{+,0,T}^{B\eta'}$ with
the parameters $f_{qs}=5.14$ MeV, $\phi=36.84^\circ$, and
$m_{qq}=0.22$ GeV selected from Fig.~\ref{figm0} are listed in
Table~\ref{tab}. The form factor values at zero recoil are larger
than those in \cite{Charng} due to the enhancement of $m_{qq}$.
Consequently, the percentages of the gluonic contribution are lower
here. It is found that the contribution from Eq.~(\ref{phiqs}) is,
like the gluonic one, unimportant. Hence, we can simply concentrate
on the effect of the modified $m_{qq}$, when studying the
$B\to\eta^{(\prime)}K^{(*)}$ decays.

\section{CRITICAL REVIEW}

As mentioned above, a large $m_{qq}$ increases the
$B\to\eta^{(\prime)}$ form factors and the
$B\to\eta^{(\prime)}\ell\nu$ branching ratios. Based on the form
factor values at maximal recoil in Table~\ref{tab} and the
parametrization for the dependence on the lepton-pair invariant mass
in \cite{Ball:2004ye}, the branching ratios can be obtained. It has
been verified that the predictions in PQCD \cite{ACG07},
\begin{eqnarray}
B(B^+\to\eta \ell^+\nu)&=&1.27\times 10^{-4}\;,\nonumber\\
B(B^+\to\eta'\ell^+\nu)&=&0.62\times 10^{-4}\;, \label{acg}
\end{eqnarray}
obey the experimental bounds \cite{BaBar}
\begin{eqnarray}
B(B^+\to\eta \ell^+\nu)&=&(0.84\pm 0.27\pm 0.21)\times
10^{-4}<1.4\times 10^{-4}\; (90\%\; {\rm C.L.})\;,\nonumber\\
B(B^+\to\eta' \ell^+\nu)&=&(0.33\pm 0.60\pm 0.30)\times
10^{-4}<1.3\times 10^{-4}\; (90\%\; {\rm C.L.})\;.
\end{eqnarray}
This check should apply to other proposals resorting to the
enhancement of the $B\to\eta^{(\prime)}$ form factors, such as the
inclusion of the flavor-singlet contribution.

Without the flavor-singlet contribution, one should have the ratio
of the $B\to\eta^{(\prime)}\ell\nu$ branching ratios,
\begin{eqnarray}
R_{\ell\nu}\equiv\frac{B(B\to \eta' \ell \nu)}{B(B\to \eta \ell
\nu)} \approx\tan^2\phi \;, \label{rp}
\end{eqnarray}
which is less than unity in the FKS scheme. The PQCD results in
Eq.~(\ref{acg}) agree with this expectation. However, the recent
CLEO measurement with $R_{\ell\nu}> 2.5$ \cite{CLEO07} may indicate
a significant flavor-singlet contribution in the
$B\to\eta^{(\prime)}$ transitions. A simple estimate shows that the
gluonic contribution must reach at least half of the quark one in
order to satisfy CLEO's bound, in conflict with the implication from
other data \cite{FKS,KP,AS02,AP03}. Furthermore, the ratio of the
observed $D_s\to\eta^{(\prime)}\ell\nu$ branching ratios
\cite{CLEO95,PDG},
\begin{eqnarray}
\frac{B(D_s\to \eta' \ell \nu)}{B(D_s\to \eta \ell \nu)}= 0.35\pm
0.09\pm 0.07\;, \label{rd}
\end{eqnarray}
does not reveal the same signal: under a monopole parametrization,
it corresponds to the ratio of the $D_s\to\eta^{(\prime)}$ form
factors at the maximal recoil \cite{BFT95},
\begin{eqnarray}
\frac{F_+^{D_s\eta'}(0)}{F_+^{D_s\eta}(0)}= 1.14\pm 0.17\pm
0.13\;,\label{fra}
\end{eqnarray}
in agreement with the expectation from the FKS scheme.

The gluonic contribution to the $B\to\eta^{(\prime)}$ transitions
also plays an essential role in the proposal of \cite{WZ0610}. It is
destructive to the quark contribution from the data fitting based on
SCET, so that the $B\to\eta^{(\prime)}$ form factors have small
values of $O(10^{-2})$. The $B\to\eta^{(\prime)}K$ branching ratios
then receive contributions mainly from the nonperturbative charming
penguin and gluonic charming penguin amplitudes. Especially, the
gluonic charming penguin is responsible for the dominance of the
$B\to\eta'K$ branching ratios over the $B\to\pi K$ ones. With the
potentially sizable gluonic contribution, the ratio $R_{\ell\nu}$ in
Eq.~(\ref{rp}) could deviate from $\tan^2\phi$. However, due to the
huge uncertainty of this contribution, no definite prediction for
$R_{\ell\nu}$ can be made. Nevertheless, it is still possible to
test the mechanism in \cite{WZ0610} by measuring the semileptonic
decays: the smallness of the $B\to\eta^{(\prime)}$ form factors
leads to the small $B\to\eta^{(\prime)}\ell\nu$ branching ratios of
$O(10^{-5})$, compared to $O(10^{-4})$ from the PQCD \cite{ACG07}
and QCDF \cite{BN02} approaches. Taking into account the uncertainty
of Solutions I and II in \cite{WZ0610} to 1$\sigma$, we estimate,
using the parametrization for form factors in \cite{Ball:2004ye},
the rough upper bounds
\begin{eqnarray}
B(B^+\to\eta \ell^+\nu)&<&5\times 10^{-5}\;,\nonumber\\
B(B^+\to\eta'\ell^+\nu)&<&3\times 10^{-5}\;. \label{wz}
\end{eqnarray}

\begin{figure}[ht]
\begin{center}
\resizebox{7cm}{!}{\includegraphics{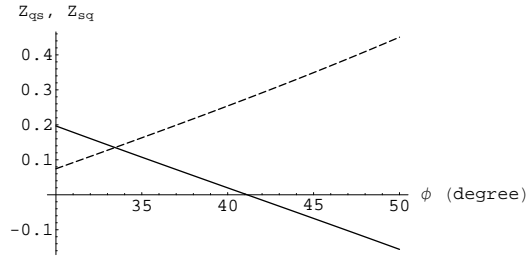}}

\caption{Dependence of $Z_{qs}$ (solid line) and $Z_{sq}$ (dashed
line) on $\phi$ in units of degrees.}\label{figm1}
\end{center}
\end{figure}

The proposal in \cite{GK06} resorts to the ratio of the matrix
elements,
\begin{eqnarray}
\left|\frac{\langle 0|\bar s \gamma_5 s|\eta'\rangle}{\langle 0|\bar
s \gamma_5 s|\eta\rangle}\right|\approx 2.1 \;,\label{mra}
\end{eqnarray}
greater than $\cot\phi \approx 1.2$ in the FKS scheme. The matrix
elements of the pseudoscalar density define the chiral mass scales,
to which the two-parton twist-3 contributions are proportional.
Therefore, the above ratio would affect Eq.~(\ref{fra}) through
these contributions in the theoretical frameworks based on the
heavy-quark expansion and factorization theorems such as PQCD and
SCET. However, the $D_s\to\eta^{(\prime)}\ell\nu$ data do not
indicate a deviation from the FKS scheme. A more convincing
discrimination can be achieved by measuring the
$B_s\to\eta^{(\prime)}\ell^+\ell^-$ decays, for which the
heavy-quark expansion works better. If the mechanism in \cite{GK06}
is valid, a significant deviation from
\begin{eqnarray}
R_{\ell\ell}\equiv\frac{B(B_s\to \eta' \ell \ell)}{B(B_s\to \eta
\ell \ell)} \approx\cot^2\phi \;, \label{rpl}
\end{eqnarray}
will be observed. According to \cite{TLS}, the twist-2 and twist-3
contributions are roughly equal in the $B_s$ meson transition form
factors. It is then likely that Eq.~(\ref{mra}) doubles the ratio
$R_{\ell\ell}$, leading to $R_{\ell\ell}\approx 3$.

On the other hand, the results in \cite{GK06} can be examined from
the viewpoint of the OZI-rule violation. The four matrix elements on
the right-hand side of the following transformation have been
derived in \cite{GK06,GK05}:
\begin{eqnarray}
\left(\begin{array}{cc} \langle 0|\bar q \gamma_5 q|\eta_q\rangle
& \langle 0|\bar s \gamma_5 s|\eta_q\rangle \\
              \langle 0|\bar q \gamma_5 q|\eta_s\rangle &
\langle 0|\bar s \gamma_5 s|\eta_s\rangle
\\
\end{array}\right)
&=&U^\dagger(\phi)\left(\begin{array}{cc} \langle 0|\bar q
\gamma_5 q|\eta\rangle
& \langle 0|\bar s \gamma_5 s|\eta\rangle \\
              \langle 0|\bar q \gamma_5 q|\eta'\rangle &
\langle 0|\bar s \gamma_5 s|\eta'\rangle
\\
\end{array}\right)\;.\label{mmgk}
\end{eqnarray}
The matrix elements on the left-hand side of Eq.~(\ref{mmgk}) define
the OZI violating quantities,
\begin{eqnarray}
Z_{qs}\equiv\frac{\langle 0|\bar q \gamma_5 q|\eta_s\rangle}{\langle
0|\bar q \gamma_5
q|\eta_q\rangle}=\frac{m_{qs}^2}{m_{qq}^2}\;,\;\;\;\;
Z_{sq}\equiv\frac{\langle 0|\bar s \gamma_5 s|\eta_q\rangle}{\langle
0|\bar s \gamma_5
s|\eta_s\rangle}=\frac{m_{sq}^2}{m_{ss}^2}\;,\label{zra}
\end{eqnarray}
which are related to the mass ratios via Eqs.~(\ref{mqq}) and
(\ref{mqs}). Figure~\ref{figm1} shows that either $Z_{qs}$ or
$Z_{sq}$ remains sizable no matter how $\phi$ is varied in the range
$30^\circ<\phi<50^\circ$: for $\phi\approx 39.3^\circ$ \cite{FKS}
($32.7^\circ$ adopted in \cite{GK06}), we have $Z_{qs}\approx 3\%$
(15\%) and $Z_{sq}\approx 24\%$ (12\%). That is, the proposal in
\cite{GK06} demands more significant OZI-rule violation, compared to
the few-percent violation in the decay constants considered in
Sec.~II. It is now clear that the mass $m_{qs}$ makes a smaller
impact on $m_{qq}$ than $f_{qs,sq}$ do: few-percent $Z_{qs}$ changes
$m_{qq}$ by only few percents following the formalism in
Eqs.~(\ref{m2qs})-(\ref{mqs}), while few-percent $Y_{qs,sq}$
increase $m_{qq}$ by a factor 2. The neglect of $m_{qs,sq}$ in
Eq.~(\ref{chiral}) is then justified.

\section{SUMMARY}

In this work we have surveyed various proposals for accommodating
the dramatically different data of the $B\to\eta^{\prime} K$ and
$B\to\eta K$ branching ratios in Eq.~(\ref{data}). The
flavor-singlet contribution \cite{BN02} seems to be insufficient for
stretching the difference under the experimental constraints from
other $\eta^{(\prime)}$ meson involved processes. If this
contribution was the responsible mechanism, both the ratios
$R_{\ell\nu}$ and $R_{\ell\ell}$ defined by Eqs.~(\ref{rp}) and
(\ref{rpl}), respectively, would deviate from the FKS expectations
by about a factor 2. Hence, it is crucial to settle down the
discrepancy between the current BaBar \cite{BaBar} and CLEO
\cite{CLEO07} measurements of the $B\to\eta^{(\prime)}\ell\nu$
decays. The dominance of the charming penguin and gluonic charming
penguin \cite{WZ0610} implies the small $B\to\eta^{(\prime)}$ form
factors and the $B\to\eta^{(\prime)}\ell\nu$ branching ratios of
$O(10^{-5})$, which can be confronted with future data. The very
different matrix elements $\langle 0|\bar{s}\gamma_{5}s|\eta\rangle$
and $\langle 0|\bar{s}\gamma_{5}s|\eta'\rangle$ caused by the axial
$U(1)$ anomaly \cite{GK06} demand larger OZI-rule violation, and
render $R_{\ell\ell}$ become twice of $\cot^2\phi$. The enhancement
of the chiral scale associated with the $\eta_q$ meson \cite{ACG07}
requires only few-percent OZI-rule violation, and both
Eqs.~(\ref{rp}) and (\ref{rpl}) hold. In summary, precise data of
the $B\to\eta^{(\prime)}\ell\nu$ and
$B_s\to\eta^{(\prime)}\ell^+\ell^-$ decays will help discriminating
the above proposals.

\vskip 1.0cm We thank M. Beneke, W.C. Chang, C.H. Chen, R.
Escribano, T. Feldmann, J.M. Frere, J.M. Gerard, E. Kou, and S.
Mishima for useful discussions. This work was supported by the
National Science Council of R.O.C. under Grant No.
NSC-95-2112-M-050-MY3 and by the National Center for Theoretical
Sciences of R.O.C.. HNL thanks Hokkaido University for the
hospitality during his visit, where this work was initiated.

\end{document}